\documentclass[runningheads]{llncs}
\usepackage{float}
\usepackage{caption}
\usepackage{subcaption}
\usepackage[T1]{fontenc}
\usepackage{bbding}
\usepackage{hyperref}
\usepackage{graphicx}
\usepackage{amssymb}
\usepackage{orcidlink}

\begin{document}
\title{OpenLifelogQA: An Open-Ended Multimodal Lifelog Question-Answering Dataset}

%
%

\author{Quang-Linh Tran\inst{1}\Envelope\orcidlink{0000-0002-5409-0916} \and
Hoang-Bao Le\inst{1}\orcidlink{0009-0000-2496-4347} \and 
Tuong-Nghiem Diep\inst{2}\orcidlink{0000-0001-7406-1250} \and \\ 
Binh Nguyen\inst{2}\orcidlink{0000-0001-5249-9702} \and
Gareth J. F. Jones\inst{1}\orcidlink{0000-0003-2923-8365} \and
Cathal Gurrin\inst{1}\Envelope\orcidlink{0000-0003-2903-3968}}

\authorrunning{Quang-Linh Tran et al.} 
\titlerunning{OpenLifelogQA Dataset}

\institute{
ADAPT Centre, School of Computing, Dublin City University, Dublin, Ireland
\email{\{quang-linh.tran2, bao.le2\}@mail.dcu.ie  \\
\{gareth.jones, cathal.gurrin\}@dcu.ie 
} \and
University of Science, Vietnam National University, Ho Chi Minh City, Vietnam
\email{dtnghiem21@apcs.fitus.edu.vn, ngtbinh@hcmus.edu.vn}
}

\maketitle              
\begin{abstract}

 We introduce \textbf{OpenLifelogQA}, a large-scale open-ended lifelog QA dataset constructed from 18 months of multimodal lifelog data. Lifelogging is the passive collection and analysis of personal daily activities using wearable devices, producing rich multimodal data such as images, locations, and biometrics. Question answering (QA) over lifelog data enables users to interactively query their own experiences, supporting applications in memory support, lifestyle analysis, and personal assistance. OpenLifelogQA contains 14,187 Q\&A pairs spanning multiple question types and difficulty levels, designed to support robust evaluation in realistic settings. Compared with prior resources, OpenLifelogQA offers greater diversity and practicality for real-world applications. To establish baselines, we evaluate the LLaVA-NeXT-Interleave 7B model, achieving 89.7\% BERTScore, 25.87\% ROUGE-L, and an average LLM Score of 3.97. By releasing OpenLifelogQA, we aim to promote future research on lifelog technologies, paving the way for personal lifelog assistants capable of memory augmentation, healthcare support, and lifestyle coaching.

\end{abstract}
\section{Introduction}

Lifelogging refers to the passive collection, storage, and analysis of personal daily life data using wearable devices \cite{cathal_lifelog}. Such data typically includes point-of-view (PoV) images from wearable cameras, biometrics, location traces, and other digital records such as emails or internet activity. Lifelog data has been widely studied for applications including memory archiving and enhancement \cite{kikhia2010building,memoriease}, health monitoring \cite{kim_lifelog_system_health}, and lifestyle analysis \cite{kumar2014towards} \cite{rossetto2025castle}. Among these, lifelog retrieval—finding specific moments in lifelog data—has been an active research topic for many years \cite{lsc24,memoriease2,memoriease3}, helping users recover forgotten events. However, retrieval alone is insufficient to satisfy users’ information needs, since many insights remain hidden within the retrieved data. This motivates the development of resources to enable question answering (QA) over lifelog data.

Lifelog QA aims to provide precise answers to user questions based on lifelog data. Unlike retrieval, which only identifies relevant content, QA must extract and reason over specific details within that content. For example, the question “What did I have for lunch on May 01, 2025, and for how long?” requires first identifying lunch-related activities, then analyzing associated images and metadata to determine both the food consumed and its duration. This task poses several challenges: (i) the long-term, large-scale nature of lifelog data requires robust retrieval to locate relevant events; (ii) many queries demand complex, multi-hop reasoning; and (iii) the inherently multimodal nature of lifelog data—images, timestamps, and locations—requires models to interpret heterogeneous information effectively. Despite these challenges, lifelog QA has enormous potential. It could allow users to query their lifelogs for purposes such as fact verification, lifestyle analysis, or memory enhancement \cite{tran2024interactive}. For example, “Did I take any medicine yesterday?” is a practical reminder-type query, while “Tell me what I discussed with J. during our last meeting” points toward future context-aware assistants capable of richer personal interactions.

Given these potential benefits, lifelog QA promises to significantly enhance how users interact with their personal data. However, progress has been constrained by the lack of high-quality datasets. Prior work has produced either small-scale collections \cite{tran2022llqa} or large but synthetic lifelog QA datasets \cite{tan-etal-2023-timelineqa}, both of which limit diversity and realism for real-world applications. Moreover, with the rise of generative AI, open-ended QA—where answers are expressed naturally rather than as spans or multiple-choice—offers a more realistic interaction paradigm. To address these gaps, this paper introduces a novel open-ended lifelog QA dataset, built on a large-scale, realistic collection from the Lifelog Search Challenge \cite{lsc24}. Our dataset contains diverse question types reflecting practical use cases, laying the foundation for robust evaluation and the development of models that can more effectively understand and interact with real-world lifelog data.


\section{Related Work}

Before discussing datasets specifically in the field of lifelog QA, we outline details of some datasets in Egocentric Video QA. Video QA shares some characteristics with lifelog QA in using PoV data and focusing on life-related questioning. QaEgo4D \cite{barmann2022did} is the largest human-curated Egocentric Video QA dataset with 1,325 videos and 14,507 QA pairs. This dataset focuses on videos of long duration and is constructed based on the Ego4D dataset \cite{grauman2022ego4d}. There are other short video duration datasets such as OpenEQA \cite{majumdar2024openeqa}, EgoVQA \cite{fan2019egovqa}, and EgoTaskQA \cite{jia2022egotaskqa}. However, even the longest video in these datasets cannot be compared to the length of a set of daily lifelog images, if these images were assembled into a video. In addition, lifelog data includes not only images, but also other data such as time, location, and biometrics, so lifelog QA can be both visual-related and metadata-related. These are the key differences between Egocentric Video QA and Lifelog QA. 

In the domain of lifelog QA, three notable datasets have been developed, each with distinct approaches and limitations. The LLQA dataset \cite{tran2022llqa} created a lifelog QA task by augmenting the Lifelog Search Challenge 2020 (LSC'20) dataset \cite{gurrin2020introduction} with around 15,000 semi-automatically generated QA pairs. This dataset had only 85 days of lifelog data and was constructed on a closed QA format with two main types of questions (multiple-choice and yes/no). Meanwhile, TimelineQA \cite{tan-etal-2023-timelineqa} takes a template-based approach to synthetically generate a massive dataset of 128 million lifelog entries. The approach first creates an imaginary persona with key information such as gender, profession, and then generates life episodes (events) with different time densities, such as daily, weekly, monthly and yearly. Finally, the QA pairs are generated based on the information of episodes. While impressive in scale, its synthetic and text-only nature reduces its realism and limits its applicability to real-world use cases and its value for research. The third dataset is MemoriQA \cite{tran2024memoriqa}, an open-ended lifelog QA dataset combining manual and synthetic QA generation. This dataset is built on 61 days of lifelog data and contains 3,644 QA pairs. This is a relatively small dataset, but it is built on a practical approach. Across all three datasets, several challenges remain from heavy dependence on synthetic data to small-sized data due to the high cost of manual annotation. This challenge restricts the diversity and practicality of the QA pairs. 

From the previous datasets, we observe a key drawback in the lack of practicality and diversity in lifelog QA. In our current work, we have constructed a broad, diverse lifelog QA dataset based on an existing large lifelog dataset from Lifelog Search Challenge \cite{lsc24} to ensure diversity and practicality with open-ended Q\&A on different types of questions. In addition, with the rise of large language models (LLMs) in recent years, generative and open-ended QA can help to increase the human-likeness of answers. We use them as a parallel annotator to human annotators to increase the efficiency of Q\&A generation. 

\section{Dataset Construction}

In this section, we provide information about the process of constructing our dataset. We start with the source of lifelog data and then the annotation process. Finally, we describe the quality control process to measure the quality of Q\&A generation.

\subsection{Data Source}

We use the lifelog dataset from Lifelog Search Challenge 2024 (LSC'24) \cite{lsc24} as the starting point to create the QA dataset. It contains 722,606 PoV images from a single anonymous lifelogger, ranging over a period of 18 months from January 2019 to June 2020. The images are captured from a PoV camera\footnote{https://getnarrative.com/}  worn on the chest of the lifelogger. The images have a resolution of 1,024 x 768 pixels. All faces and readable text have been removed, and certain scenes and activities have been manually filtered out to respect local privacy requirements. In addition to the images, we also use the metadata of the lifelog collection. This includes the capture time of each image, the capture location of each image, and every minute if there is no image captured at that time, and related biometrics from a smartwatch. From this data source, we perform the QA generation described in the next section.

\subsection{QA Generation Process}

\begin{figure*}
    \centering
    \includegraphics[width=1\linewidth]{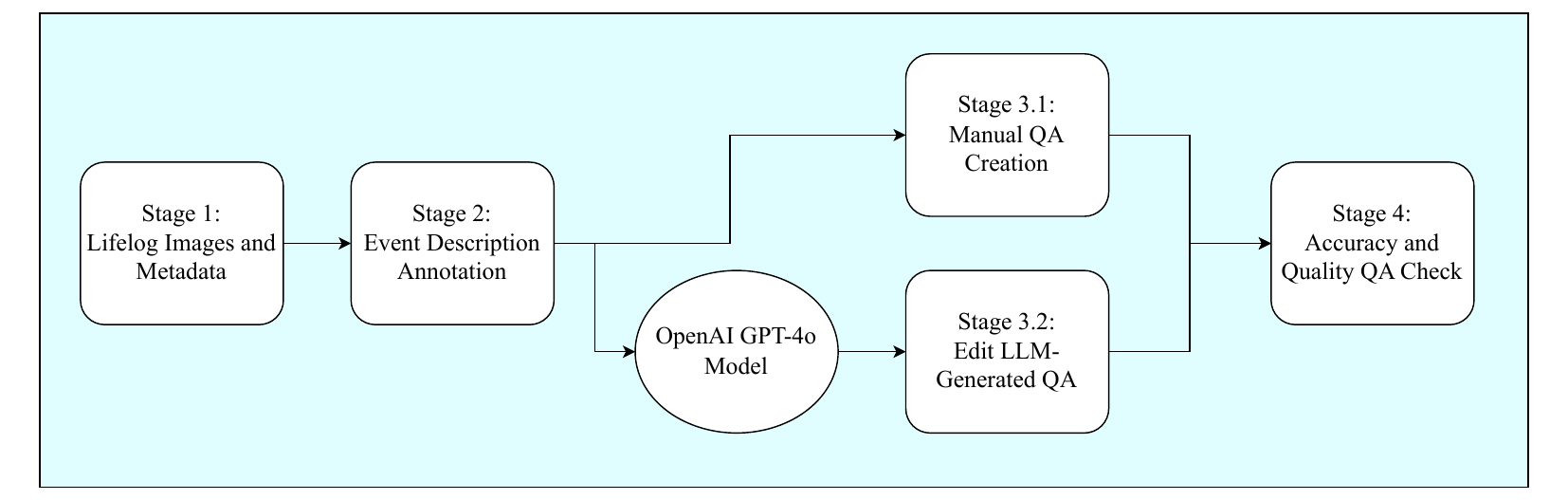}
    \caption{Annotation Process}
    \label{fig:annotation_process}
\end{figure*}

From the lifelog data source, and in line with prior lifelog QA datasets, we first generated \textit{event descriptions} to serve as the basis for questions and answers. To support this process, we recruited 10 undergraduate volunteers majoring in Data Science and Information Technology. Their English proficiency ranged from good to proficient (TOEIC $\geq$ 650), and they possessed basic knowledge of QA tasks. The volunteers participated in two training sessions covering lifelog concepts, lifelog QA, and the annotation workflow. In addition, a comprehensive guideline was provided to describe the overall procedure and address common annotation issues.  

We performed QA generation month by month, beginning with January 2019. Each volunteer was assigned four days of lifelog data per month, with one day overlapping among all volunteers to assess the consistency of event description generation. The process consisted of four stages, as illustrated in Figure~\ref{fig:annotation_process}. An annotation platform was developed to support these tasks.  

\begin{itemize}
    \item \textbf{Stage 1: Event Review.} Volunteers reviewed all lifelog images and associated metadata for a given day to obtain an overview of the lifelogger’s activities.  

    \item \textbf{Stage 2: Event Description.} Volunteers segmented the data into events and wrote first-person descriptions for each. An event was defined as a sequence of images representing a single activity (e.g., eating, watching TV, or working on a laptop). Each event was associated with a start time and a representative image, while the end time was defined by the start of the next event. For example: \textit{``I am working on my laptop''} at 2019-01-01 12:23:42, with ImageID \texttt{20190101\_122342\_000}.  

    \item \textbf{Stage 3: QA Generation.} Both volunteers and the GPT-4o model \cite{hurst2024gpt} generated QA pairs from event descriptions. Each volunteer created 10 QA pairs per day, while GPT-4o generated 20 QA pairs, balancing quality, scalability, and human effort. Every QA pair was linked to one or more events from Stage 2, with corresponding event IDs (the ImageID of the event’s start time) recorded as ground truth for retrieval tasks. The QA format was deliberately unconstrained to encourage diversity and practicality. We particularly emphasised complex aggregation questions requiring reasoning or calculation (e.g., counting or measuring duration). For instance: \textit{``How long did I drive from my house to the church on the morning of January 21, 2020?''}.  

    \item \textbf{Stage 4: Quality Control.} Volunteers conducted a final review to verify the accuracy of QA pairs and event descriptions. One day per month was annotated by all volunteers to measure consistency in event description creation. We measure the BERT Score \cite{bertscore} between event descriptions generated by annotators and ensure a threshold of 0.85 BERT Score. This score indicates a high semantic relevance and agreement of annotators on event descriptions. Since every annotator is freely generating QA, so we cannot measure the inner annotator agreement. Instead, volunteers cross-reviewed and corrected GPT-4o outputs, resolving errors, duplicates, and inconsistencies before finalizing them as official QA. There are about 200 QAs are corrected after the cross-check.
\end{itemize}


\section{Dataset Information}

In this section, we provide information about the OpenLifelogQA dataset after the QA generation process. We describe the information and statistics of the dataset, as well as some analysis of questions and answers. Finally, we discuss licensing and ethical \& privacy considerations arising from this, and any similar lifelog dataset.

\begin{table*}[t]
\caption{Statistics on three sets of OpenLifelogQA.}
\resizebox{\textwidth}{!}{
\begin{tabular}{|c|r|r|r|r|r|r|}
\hline
\textbf{Set} &
  \multicolumn{1}{c|}{\begin{tabular}[c]{@{}c@{}}\# \textbf{atomic}\\ \textbf{question}\end{tabular}} &
  \multicolumn{1}{c|}{\begin{tabular}[c]{@{}c@{}}\# \textbf{temporal}\\ \textbf{question}\end{tabular}} &
  \multicolumn{1}{c|}{\begin{tabular}[c]{@{}c@{}}\# \textbf{aggregation}\\ \textbf{question}\end{tabular}} &
  \multicolumn{1}{c|}{\begin{tabular}[c]{@{}c@{}}\# \textbf{context}\\ \textbf{events}\end{tabular}} &
  \multicolumn{1}{c|}{\begin{tabular}[c]{@{}c@{}}\textbf{Avg} \textbf{question}\\ \textbf{length}\end{tabular}} &
  \multicolumn{1}{c|}{\begin{tabular}[c]{@{}c@{}}\textbf{Avg} \textbf{answer}\\ \textbf{length}\end{tabular}} \\ \hline
Training   & 6578 & 946 & 3653 & 1.64 & 12.74 & 6.67 \\ \hline
Validation & 895  & 131  & 476  & 1.75 & 11.99 & 5.74 \\ \hline
Testing    & 882  & 156  & 470  & 1.75 & 12.23 & 5.73 \\ \hline
\end{tabular}
}
\label{tab:dataset_stat}
\end{table*}

\begin{table*}[t]
\caption{Lifelog QA datasets comparison.}
\centering
\resizebox{\textwidth}{!}{
\begin{tabular}{|c|r|c|c|r|c|c|c|c|}
\hline
\textbf{Dataset} &
  \multicolumn{1}{c|}{\# \textbf{QA}} &
  \begin{tabular}[c]{@{}c@{}}\textbf{Lifelog}\\ \textbf{data}\end{tabular} &
  \begin{tabular}[c]{@{}c@{}}\textbf{Question}\\ \textbf{format}\end{tabular} &
  \multicolumn{1}{c|}{\begin{tabular}[c]{@{}c@{}}\# \textbf{Event}\\ \textbf{description}\end{tabular}} &
  \textbf{Image} &
  \begin{tabular}[c]{@{}c@{}}\textbf{Time} \&\\ \textbf{Location}\end{tabular} &
  \textbf{Open-sourced} \\ \hline
LLQA \cite{tran2022llqa} &
  15,065 &
  85 days &
  Multiple-choice &
  \multicolumn{1}{c|}{None} &
  \checkmark &
  \checkmark &
  \checkmark \\ \hline
Timeline QA \cite{tan-etal-2023-timelineqa} &
  600,000 &
  Synthetic &
  Open-ended &
  128,023,476 &
   &
  \checkmark &
  \checkmark \\ \hline
MemoriQA \cite{tran2024memoriqa} &
  3,644 &
  61 days &
  Open-ended &
  1,925 &
  \checkmark &
  \checkmark &
   \\ \hline
OpenLifelogQA &
  14,187 &
  514 days &
  Open-ended &
  27,705 &
  \checkmark &
  \checkmark &
  \checkmark \\ \hline
\end{tabular}
}
\label{tab:dataset_comparision}
\end{table*}

\subsection{Descriptions}

After the QA generation process, we obtained a lifelog QA dataset with 27,705 event descriptions and 14,187 QA pairs. This data spans 514 days of lifelog data, which means an average of 54 events and 28 questions per day. The events and QA pairs are independent, as a QA pair can involve between 1 and several events, and some events can relate to no QA. The average number of events for each QA pair is 1.66 events. Table \ref{tab:examples} shows some examples of QA pairs and groundtruth contexts of the OpenLifelogQA dataset. We also compare this dataset with previous datasets in table \ref{tab:dataset_comparision}. Our proposed OpenLifelogQA dataset offers a more comprehensive resource for lifelog QA research. This dataset includes a richer set of events and more diverse lifelog data than previous datasets, except for the synthetic data in Timeline QA. It also incorporates open-ended questions, as well as images, time, and location metadata. 

\begin{table*}[t]
\caption{Some QA pairs and ground-truth context examples}
\resizebox{\textwidth}{!}{
\begin{tabular}{|l|l|l|l|}
\hline
\multicolumn{1}{|c|}{\textbf{Question}} & \multicolumn{1}{c|}{\textbf{Answer}} & \multicolumn{1}{c|}{\textbf{Context}} & \multicolumn{1}{c|}{\textbf{Type}} \\ \hline
\begin{tabular}[c]{@{}l@{}}How long did I spend at the store\\ on the evening of June 26, 2020?\end{tabular} & \begin{tabular}[c]{@{}l@{}}I spent 23 minutes\\ at the store.\end{tabular} & \begin{tabular}[c]{@{}l@{}}20200626\_203417\_000: I am at a store,\\ 20200626\_205727\_000: I am driving away\\ from the store\end{tabular} & Aggregation \\ \hline
\begin{tabular}[c]{@{}l@{}}How many times did I watch TV\\ on June 1, 2020?\end{tabular} & \begin{tabular}[c]{@{}l@{}}I watched TV 3\\ times.\end{tabular} & \begin{tabular}[c]{@{}l@{}}20200601\_115751\_000: I am watching TV.\\ 20200601\_165938\_000: After dinner, I \\ started watching TV.\\ 20200601\_182032\_000: I stopped working\\ and started watching TV again.\end{tabular} & Aggregation \\ \hline
\begin{tabular}[c]{@{}l@{}}When did I go to the kitchen to\\ make a smoothie on June 21, 2020?\end{tabular} & \begin{tabular}[c]{@{}l@{}}I went at \\ 11:02 AM.\end{tabular} & \begin{tabular}[c]{@{}l@{}}20200621\_110203\_000: I go to the kitchen\\ to make a smoothie\end{tabular} & Atomic \\ \hline
\begin{tabular}[c]{@{}l@{}}What did I do before preparing \\ to leave the house on June 21, 2020?\end{tabular} & I watched TV. & \begin{tabular}[c]{@{}l@{}}20200621\_061246\_000: I continue to sit \\ and watch TV\\ 20200621\_062502\_000:  I have left the\\ house and am moving by car\end{tabular} & Temporal \\ \hline
\end{tabular}
}
\label{tab:examples}
\end{table*}

We split the dataset into three sets for model development and evaluation, including training, validation and test sets. The splits were determined by the month of lifelog data, April and May 2020 is used for validation, March and June 2020 for testing, and the rest for training. This splitting strategy aims to avoid data leakage in model development and evaluation. Statistics of the three datasets are shown in Table \ref{tab:dataset_stat}. The training set contains 11,159 QA pairs, while the validation and testing sets each include approximately 1,500 QA pairs. An average question has about 12 words, and answers 6 words. This length suggests concise, direct questions and answers in real-world applications. We also categorise the questions into several types, which are described in the following section.

\subsection{Analysis}

The questions are divided into three categories: atomic, temporal, and aggregation, following classifications used in prior studies \cite{tan-etal-2023-timelineqa}. 

\begin{itemize}
    \item \textbf{Atomic questions} can be answered using a single context, such as ``\textit{What time did I have lunch yesterday?}''. 
    \item \textbf{Temporal questions} involve reasoning about the sequence of events in lifelog data, asking about what happened before or after a known event, such as ``\textit{What did I do after having lunch yesterday?}''.
    \item \textbf{Aggregation questions} require synthesizing information across multiple events to derive an answer, such as calculating a duration, count, or total. An example is: ``\textit{How long did I have lunch yesterday?}''. 
\end{itemize}

\begin{figure*}
    \centering
    \begin{subfigure}[b]{0.45\linewidth}
    \centering
        \includegraphics[width=1\linewidth]{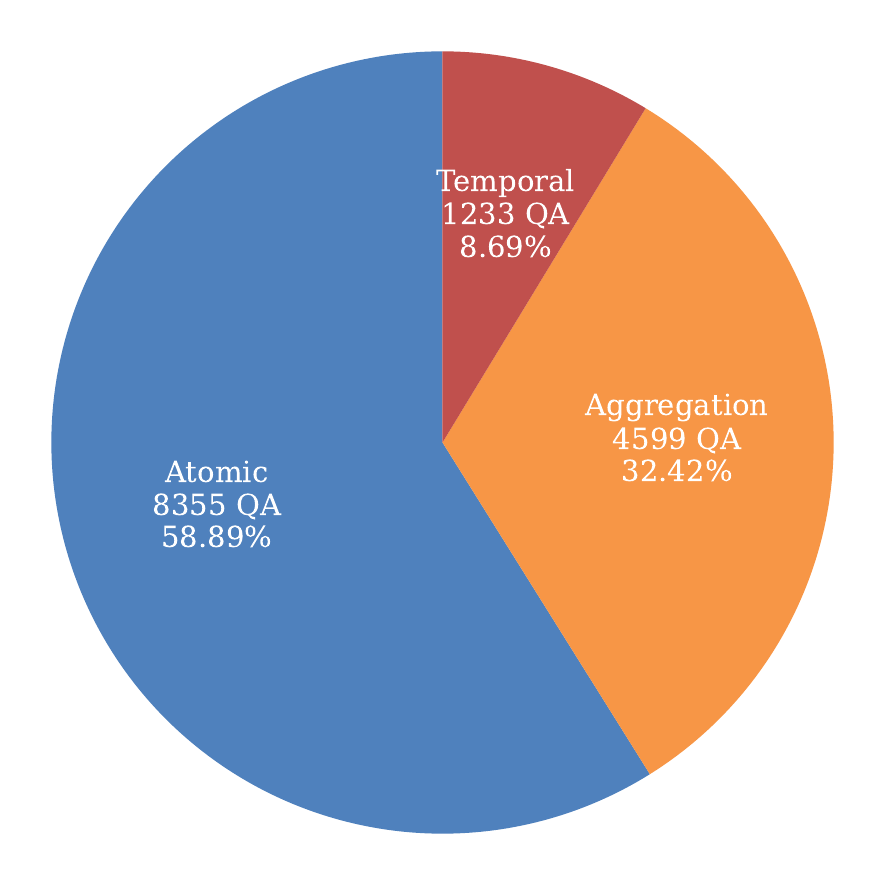}
        \caption{Question type distribution}
        \label{fig:question_type}
    \end{subfigure}
    \hfill
    \begin{subfigure}[b]{0.42\linewidth}
    \centering
        \includegraphics[width=1\linewidth]{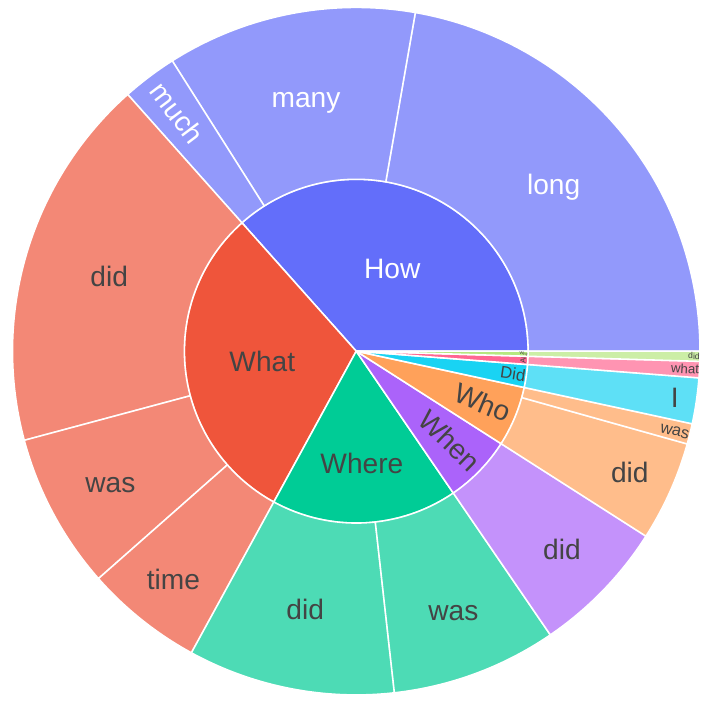}
         \caption{Question word distribution}
        \label{fig:question_word}
    \end{subfigure}
    \caption{Analysis on questions}
    \label{fig:analysis}
\end{figure*}

The distribution of these types of questions is illustrated in Figure \ref{fig:question_type}. We further analyse the distribution of question words in Figure \ref{fig:question_word}. The distribution of question words shows that the dataset reflects realistic lifelog use cases, with a strong focus on ``what'', ``where'', ``when'' and ``how'' questions, and less attention to more exploratory or uncommon queries. 

\subsection{Licensing and Access}

For the purpose of fostering research in lifelog QA, we release the dataset for research purposes only. The dataset is available under a Creative Commons Attribution-NonCommercial 4.0 International License. Anyone interested in this dataset can contact the authors of this paper to complete a user agreement form to obtain the event descriptions and QA pairs of the OpenLifelogQA dataset. The images and other metadata from the LSC lifelog dataset are provided separately at the main page of Lifelog Search Challenge \footnote{\href{http://lifelogsearch.org/lsc/lsc_data/}{http://lifelogsearch.org/lsc/lsc\_data/}}. This is to ensure the privacy and ethical issues in lifelog data. For the reviewers of this dataset paper, we provide a link to the dataset to review the content of the dataset as follows: \href{https://tinyurl.com/OpenlifelogQA}{Link}.

\subsection{Ethical Considerations and Privacy}

As lifelog data is private and prone to many privacy problems due to the highly sensitive nature of personal data, which can include health details, social interactions, and location history, the use of this dataset should be considered in an ethical and privacy context. The data source of LSC'24 has already removed all faces and readable text in the lifelog images to ensure privacy for bystanders. Moreover, any analysis or model training on lifelog data should avoid generating intrusive inferences that could compromise individual autonomy or dignity. Ethical oversight and compliance with data protection regulations such as GDPR are essential to maintain public trust and ensure responsible research practices.

\section{Baseline Experiment}

In this section, we run a baseline experiment on the OpenLifelogQA dataset with a well-known and easy-to-reproduce multi-modal LLM, LLaVA-NeXT-Interleave. The experimental setting and results, as well as some analysis, are discussed in the following sections.

\subsection{Settings}

From the test set of OpenLifelogQA, we use the LLaVA-NeXT-Interleave model \cite{li2024llavanextinterleavetacklingmultiimagevideo} to predict the answers to questions without finetuning. This is to measure the performance of a multi-modal LLM on lifelog data. The LLaVA-NeXT-Interleave model is a multi-modal LLM with the Qwen \cite{bai2023qwentechnicalreport} as the base LLM model and Siglip \cite{zhai2023sigmoidlosslanguageimage} as the base vision tower. This model is fine-tuned with visual instructions on multi-image, video, 3D, and multi-patch of single images. The model shows leading results in various benchmarks as well as several emerging capabilities. With a strong performance on various tasks, we use this model as the baseline to run on the OpenLifelogQA dataset. We use two versions of the model, which are Qwen 7B and 0.5B, to see the effect of model size on the performance on a test set of OpenLifelogQA. We do not fine-tune the model on our dataset due to the limited computing resources, as well as to measure the performance of transferring tasks from multi-modal data to lifelog data of LLaVA-NeXT-Interleave.

\begin{table*}[t]
\caption{Predicted answers versus ground-truth answers and LLM Score.}
\resizebox{\textwidth}{!}{
\begin{tabular}{|l|l|l|r|}
\hline
\multicolumn{1}{|c|}{\textbf{Question}} & \multicolumn{1}{c|}{\textbf{Ground-truth Answer}} & \multicolumn{1}{c|}{\textbf{Predicted Answer}} & \multicolumn{1}{c|}{\textbf{LLM Score}} \\ \hline
\begin{tabular}[c]{@{}l@{}}How long did I spend at the store\\ on the evening of June 26, 2020?\end{tabular} & \begin{tabular}[c]{@{}l@{}}I spent 23 minutes\\ at the store.\end{tabular} & \begin{tabular}[c]{@{}l@{}}I spent 17 minutes at the store\\ on the evening of June 26, 2020.\end{tabular} & 3 \\ \hline
\begin{tabular}[c]{@{}l@{}}How many times did I watch TV\\ on June 1, 2020?\end{tabular} & I watched TV 3 times. & I watched TV three times on June 1, 2020. & 5 \\ \hline
\begin{tabular}[c]{@{}l@{}}When did I go to the kitchen to\\ make a smoothie on June 21, 2020?\end{tabular} & I went at 11:02 AM. & \begin{tabular}[c]{@{}l@{}}I went to the kitchen to make a smoothie on\\ June 21, 2020 at 11:02 AM.\end{tabular} & 5 \\ \hline
\begin{tabular}[c]{@{}l@{}}What did I do before preparing\\ to leave the house on June 21, 2020?\end{tabular} & I watched TV. & I was watching TV at home & 5 \\ \hline
\end{tabular}
}
\label{tab:predicted_example}
\end{table*}

We provide the question, groundtruth contexts, and the images of these contexts, as some examples in Figure \ref{tab:examples}, to the model and ask for an answer. We form a prompt with a simple instruction: ``Please act as a life assistant to answer questions based on the given context. You are given a question about life and a context describing time, location, and a description of an event involving a person. Your goal is to provide an answer to that question based on deriving information from the context. Only provide the answer from the given context and derive the answer carefully. The time in the context is the start time of the event, and the end time of the event is the start time of the next event. Thinking step by step to derive the answer. Provide a short answer with no explanation. Context: <image1> <context1>, ... <imageN> <contextN>. Question: <question>. Answer: ''. We have configured some hyperparameters, such as temperature = 0, top\_p = 0. All the experiments are run on a single A100 GPU. This experiment is just a baseline to evaluate the effectiveness of multi-modal LLM models for lifelog QA. In real-world applications, the contexts are not always available, and there is a need to retrieve the correct contexts for the QA model. 

We use three metrics of three evaluation approaches for Open-ended QA to evaluate the performance of the model, including lexical matching, semantic matching, and LLM-based evaluation. Lexical matching focuses on word-level overlap between predicted and ground truth answers, using metrics such as Exact Match, F1 score, and ROUGE \cite{lin-2004-rouge}, which measures n-gram and sequence similarity. Although effective, lexical metrics may overlook semantically correct answers with different phrasing. To address this, semantic matching evaluates the underlying meaning using metrics like BERT Score \cite{bertscore}, which compares BERT-based embeddings \cite{devlin2018bert} of predicted and reference answers. LLM-based evaluation \cite{llm_eval} involves using large language models like GPT-4o \cite{hurst2024gpt} to assign a score from 1 to 5 based on answer correctness. In this work, we adopt ROUGE, BERT Score, and LLM Score evaluation to comprehensively evaluate model performance.

\subsection{Results}

Table \ref{tab:result} shows the experimental results of the LLaVa-NeXT-Interleave model on the test set. We can see that the 7B model gives a higher LLM Score at 3.9665 than the 0.5B at 3.3873. However, the 0.5B provides a better performance at BERT Score and ROUGE-L at 90.63\% and 26.13\%, respectively. This can be explained since the 0.5B model provides predicted answers that match lexically and semantically with the groundtruth answers, but does not truly give the most accurate answer. For example, the predicted answer: ``I drive to school in 25 minutes'' and the groundtruth answer ``I drive to school in 5 minutes'' have a very high BERT Score and ROUGE-L due to the match of words and meaning, but the main important information 25 minutes and 5 minutes are wrong so LLM Score is low. The overall performance of the two models shows that the multi-modal LLM can solve this task with high performance when the correct contexts are provided. However, the words of predicted and groundtruth answers can be different, resulting in a low ROUGE-L score, but this can be improved using instruction tuning on the training set. 

\begin{table}[t]
\centering
\caption{Experimental results from the LLaVa-NeXT-Interleave models on the test set.}
\begin{tabular}{c|c|c|c}
\textbf{\# Parameters} & \textbf{BERT Score} & \textbf{ROUGE-L} & \textbf{LLM Score} \\ \hline
0.5B & 0.9063 & 0.2613 & 3.3873 \\
7B & 0.8970 & 0.2587 & 3.9665 
\end{tabular}
\label{tab:result}
\end{table}

\subsection{Analysis}

Table \ref{tab:predicted_example} shows some qualitative examples of predicted answers from the 7B model versus groundtruth answers and their score. We can see that the model mostly generates correct answers, but using different words and with more information than needed in the groundtruth answers. Only the first predicted answer is wrong due to the time calculation of the model, also there is an event of buying food in the store that confuses the model about the time of leaving the store. 

\begin{table}[t]
\centering
\caption{Performance of 7B model on different question types}
\begin{tabular}{c|c|c|c}
\textbf{Question type} & \textbf{BERT Score} & \textbf{ROUGE-L} & \textbf{LLM Score} \\ \hline
Atomic & 0.8938 & 0.2572 & 4.2574 \\
Temporal & 0.8935 & 0.2627 & 4.2115 \\
Aggregation & 0.9042 & 0.26 & 3.3362
\end{tabular}
\label{tab:question_type_performance}
\end{table}

We also analyse the performance of LLaVa-NeXT-Interleave 7B model on different types of questions. Results are shown in Table \ref{tab:question_type_performance}. Atomic is the easiest question to solve, with an LLM Score of 4.2574, but it has the lowest ROUGE-L due to the diversity in words in the answer. Aggregation is the most difficult question because it requires calculation or counting in the context, so the LLM Score is only 3.3362. However, it has the highest BERT Score and the second highest ROUGE-L because its answers mainly contain the questions with a number. This analysis highlights the challenge of aggregation questions in the lifelog QA problem.

\section{Conclusion}

In this paper, we introduced \textbf{OpenLifelogQA}, a large-scale open-ended QA dataset for lifelog data. The dataset contains 14,187 QA pairs and over 27K event descriptions collected from 18 months of lifelogging with more than 725K images and metadata. Unlike prior resources, OpenLifelogQA combines \textit{scale, diversity, and realism}, offering a strong foundation for practical lifelog applications. Our dataset captures a wide range of high-level daily life questions, supporting tasks such as lifestyle analysis, memory exploration, and personal data interaction.  

We also conducted baseline experiments with the LLaVA-NeXT-Interleave model, which achieved 25.87\% ROUGE-L and a 3.97 LLM Score without fine-tuning. These results underscore both the promise and the difficulty of lifelog QA, particularly for reasoning-intensive aggregation questions.  

Looking ahead, we plan to extend our work by exploring temporal reasoning and chain-of-thought methods to better address complex queries. More broadly, we envision OpenLifelogQA enabling future \textbf{personal lifelog assistants} capable of context-aware memory aids, healthcare monitoring, and lifestyle coaching. By releasing this dataset, we aim to foster research that brings lifelog technologies closer to practical, human-centered applications.

\section*{Acknowledgments}
This research was conducted with the financial support of Research Ireland at ADAPT, the Research Ireland Centre for AI-Driven Digital Content Technology at Dublin City University [13/RC/21-06\_P2] and [18/CRT/6223]. For the purpose of Open Access, the author has applied a CC BY public copyright licence to any Author Accepted Manuscript version arising from this submission.

\bibliographystyle{splncs04}
\bibliography{sample-base}

\end{document}